# TEMPERATURE-DEPENDENT MAGNETIZATION REVERSAL IN EXCHANGE BIAS NiFe/IrMn/NiFe STRUCTURES

Ch. Gritsenko[1], I. Dzhun[2], M. Volochaev[3], M. Gorshenkov[4], G. Babaytsev[2], N. Chechenin[2,5], A. Sokolov[3], Oleg A. Tretiakov[6,7], and V. Rodionova[1,4]

[1] *Immanuel Kant Baltic Federal University, A. Nevskogo 14, 236041 Kaliningrad, Russia*
[2] *Skobeltsyn Institute of Nuclear Physics, Lomonosov Moscow State University, Leninskie Gory 1/2, 119991 Moscow, Russia*
[3] *Kirensky Institute of Physics, Federal Research Center KSC SB RAS, Akademgorodok 50/38, Krasnoyarsk, 660036, Russia*
[4] *National University of Science and Technology MISIS, Leninsky Prospect 4, Moscow, 119049, Russia*
[5] *Faculty of Physics, Lomonosov Moscow State University, Leninskie Gory 1/2, 119991 Moscow, Russia*
[6] *Institute for Materials Research and Center for Science and Innovation in Spintronics, Tohoku University, Sendai 980-8577, Japan*
[7] *School of Physics, The University of New South Wales, Sydney 2052, Australia*

Abstract

We demonstrate the magnetization reversal features in NiFe/IrMn/NiFe thin-film structures with 40% and 75% relative content of Ni in Permalloy in the temperature range from 80 K to 300 K. At the descending branches of the hysteresis loops, the magnetization reversal sequence of the two ferromagnetic layers is found to depend on the type of NiFe alloy. In the samples with 75% relative content of Ni, the bottom ferromagnetic layer reverses prior to the top one. On the contrary, in the samples with 40% of Ni, the top ferromagnetic layer reverses prior to the bottom one. These tendencies of magnetization reversal are preserved in the entire range of temperatures. These distinctions can be explained by the morphological and structural differences of interfaces in the samples based on two types of Permalloy.

PACS: 75.80.+q, 96.12.Hg, 75.50.Bb

*Keywords:* magnetization reversal; exchange bias; Permalloy

## 1. Introduction

The exchange bias phenomenon can be observed in a system of adjacent antiferromagnetic (AFM) and ferromagnetic (FM) layers under the condition of an induced uniaxial magnetic anisotropy. It leads to a shift of the hysteresis loop along the field axis. Along a fixed direction, the ferromagnetic layer becomes harder for magnetization reversal due to the exchange coupling interaction with the antiferromagnetic layer. As a result, the FM-layer is considered as a pinned layer, which is widely used in spin valves [1–4], magnetic sensors [5], and MRAM [6]. Depending on the application, the most important features of magnetization process can be either the magnitude of the exchange bias or sequence of ferromagnetic layers magnetization switching, reflected in peculiarities of hysteresis loops shape [7]. The AFM-layer thickness plays a crucial role for the exchange bias effect. Therefore, it is important to study how the material parameters and characteristic properties of the exchange bias systems influence the aforementioned features.

The exchange bias phenomenon has been studied for more than a half of a century [8], and it has been confirmed to have a strong dependence on the types of FM and AFM materials [9]. Commonly used in exchange bias systems ferromagnetic materials are Ni, Co, Fe, their alloys, and the alloys doped with impurities of elements, such as Pt. Depending on magnetization, coercive force, and magnetic anisotropy of these FM materials, one can find different exchange bias values and coercivity for a required application [10–13]. NiFe alloys have small coercivity, high initial and maximum magnetic permeabilities, as well as corrosion resistance that can be useful for digital memory devices [14,15]. There are two types of Permalloy: 'High-nickel' Permalloy [16,17], which contains 72%-80% of Ni, and 'low-nickel' Permalloy [17], which contains 40%-50% of Ni. The 'High-nickel' Permalloy has a small crystalline anisotropy, large initial permeability, and is usually used in traditional exchange bias systems [18–20]. The 'Low-nickel' Permalloy has higher crystalline anisotropy and larger saturation magnetization in comparison with the 'high-nickel' Permalloy. The 'Low-nickel' one is usually used in write heads [15].

One of the ways to improve the required properties of the exchange biased systems is to use the trilayer structures instead of bilayers [9,21], where the two ferromagnetic layers are separated by the antiferromagnetic one. Such structures produce step-wise hysteresis loops [22,23] due to the two exchange-coupled interfaces with different energies.

Exchange bias effect depends strongly on temperature [24,25]. The thicknesses of both ferromagnetic and antiferromagnetic layers also affect the temperature dependence of the exchange bias. In particular, for the NiFe/FeMn/NiFe systems a peak-like behavior of the exchange bias was observed in the temperature dependences for thicknesses of the seed NiFe layer from 10 nm to 50 nm, while for the 5 nm thickness the temperature dependence had been shown to exhibit a different behavior [26]. Such a behavior has been found to depend on the bottom FM-layer thickness of FM/AFM/FM trilayer systems and is related to the amount of FM-layer spins that are strongly coupled with the AFM. Concerning the blocking temperature of exchange bias (the temperature at which the exchange bias becomes nonzero), it was reported in [27] that this temperature decreases with decreasing antiferromagnetic layer thickness in both NiFe/IrMn and IrMn/NiFe structures, as well as the same dependence was observed for different CoO layer thicknesses in CoO/Co/Ge films [28]. In [25], it has been shown that as temperature changes, the domain structure of NiFe/IrMn, in particular the size of domain walls, also changes affecting the exchange bias.

In this work, the FM/AFM/FM trilayer compositions with either low- or high-nickel Permalloy have been studied. The magnetization reversal features as well as exchange bias and coercivity have been found to depend on temperature in the range from 80 K to 300 K.

## 2. Experimental details

The NiFe/IrMn/NiFe thin-film structures were fabricated by magnetron sputtering at an ambient temperature in Ar atmosphere with pressure of 3 mTorr. Magnetic field of 500 Oe was applied in plane of the substrate during the deposition process to induce the unidirectional anisotropy in samples. The substrate was Si/SiO2 (100). The buffer Ta layer with a thickness of 30 nm was deposited onto the substrate to improve the growth of further layers. For each structure, the FM-layer that was deposited prior to the other one (i.e. onto the Ta buffer layer) is denoted as "bottom". Accordingly, the FM layer that was deposited on top of the IrMn layer is denoted as "top". The last layer of 30 nm of Ta was deposited on top of each sample to prevent them from the oxidation. We prepared two series of samples, one using 'low-nickel' Permalloy that is $Ni_{40}Fe_{60}$ (LNiPy), and the other using 'high-nickel' Permalloy that is $Ni_{75}Fe_{25}$ (HNiPy). Two targets of separated Ni and Fe were used for co-deposition of NiFe alloys. The target $Ir_{45}Mn_{55}$ alloy was used for deposition of an antiferromagnetic layer. The layers thicknesses were set by the deposition time with the deposition rates estimated from the measurements of the thickness of the calibration samples by the Rutherford backscattering method. The NiFe layers were fabricated to have thickness of 10 nm, while the thickness of IrMn layers was varied to be 2, 4 or 10 nm.

The study of the samples structural properties was carried out using transmission electron microscopy (TEM) using Hitachi HT7700 microscope at the accelerating voltage of 100 kV. Cross-section pieces of samples were prepared using a Hitachi FB2100 (FIB) single-beam focused ion beam system. The magnetic properties of the samples were investigated using a Vibrating Samples Magnetometer (VSM, Lake Shore, Model 7400). The hysteresis loops for each sample were measured for in-plane geometry with the magnetic field of the VSM oriented along the induced unidirectional anisotropy, in the temperature range from 80 K to 300 K.

## 3. Results and discussion

The TEM cross-sectional images, presented in Fig. 1, show that the interfaces between the NiFe layers and IrMn are the smooth except the interface between the bottom LNiPy and IrMn. In the case of the LNiPy/IrMn interface, the partial intermixing of layers is observed. It can occur because the LNiPy grows with a large grain size [29,30] due to the stoichiometric ratio of elements in the alloy [16]. The large grain size of LNiPy causes the enhanced roughness of the LNiPy layer surface.

Assuming the IrMn layer grains to be smaller than LNiPy grains [31,32], it can possibly fill the gaps between the LNiPy grains.

Figure 2 (a) shows the hysteresis loops for the samples NiFe/IrMn(10 nm)/NiFe with LNiPy (red curve) and HNiPy (black curve). As it can be seen, the loops exhibit the step-like shape that corresponds to the separate magnetization reversal of the two ferromagnetic layers. This is more visible in Fig. 2 (b), (c), which shows the differential susceptibility (i.e. the first derivative of magnetic moment by magnetic field) distributions for the descending and ascending branches of the hysteresis loops for the aforementioned samples. The steps in the black hysteresis loop are of equal height because the two FM layers of HNiPy are of the same thickness 10 nm, as it was confirmed by the TEM (Fig. 1). However, the red hysteresis loop has non-equal heights of the steps in the descending branch: the bottom step is of smaller height than the top one. This may be caused by the partial intermixing of the bottom LNiPy layer with IrMn layer demonstrated earlier [29,33].

It should be mentioned that in the HNiPy sample in the descending branch of the hysteresis loop, the bottom FM-layer reverses prior to the top one. This can be confirmed by the coercivities for the top and bottom subloops that correlate perfectly with the coercivities of the separate HNiPy/IrMn and IrMn/HNiPy structures [34]. However, for the LNiPy sample in the descending branch of the hysteresis loop, the top FM-layer reverses prior to the bottom one. This can be revealed from the observed difference in the heights of the steps in the hysteresis loops as it was described above. Thus, the difference in magnetization reversal for the HNiPy and LNiPy samples is observed.

The typical hysteresis loops for the LNiPy and HNiPy NiFe/IrMn/NiFe samples at temperatures 90 K, 200 K, and 300 K, with thicknesses of AFM-layer of 2 nm and 4 nm, are presented in Fig. 3. The remnant magnetization of the samples at 2 nm of IrMn layer (Fig. (a), (b)) increases with the decrease of temperature. For all samples, when the temperature decreases, the coercive force increases. These facts can be explained in terms of thermal fluctuations model [35], according to which the spin structure at the interfaces becomes more stable, because when temperature is decreased it reduces the thermal-fluctuations energy of AFM atoms and therefore of the AFM-grains.

For the LNiPy samples with 2 nm of AFM-layer the kinks appear in the sections of the loops preceding the saturation state (Fig. 3 (a)). This is more visible in the differential susceptibility distribution (Fig. 4 (a)). The second peak is observed in the descending branch. This may mean that due to an increase in the anisotropy energy of the sample layers, some of the magnetic moments of the FM-layer reverse later, after the majority of the moments is already reversed. At the same time, an increase in the slope of the hysteresis loops for the LNiPy samples indicates a change in the mechanism of the magnetization reversal of the samples. Thus, if at room temperature, the contribution of magnetization reversal through the motion of the domain wall was greater than through the rotation of the magnetic moments, then with a decrease of temperature, the contribution to the process of magnetization reversal through the motion of the domain wall decreases.

At 4 nm of IrMn layer thickness the hysteresis loops for the LNiPy samples are asymmetric, since in the descending branch the two ferromagnetic layers reverse at different values of the magnetic field, whereas in the ascending branch it occurs at the same value of the field. The sequence of the magnetization reversal for two LNiPy layers is the same as at 10 nm of AFM-layer thickness at room temperature (Fig. 2 (a)). For the HNiPy samples, the magnetization-reversal sequence of the top and bottom FM-layers is the same as found for the HNiPy/IrMn/HNiPy sample with 10 nm of AFM layer at room temperature (Fig. 2 (a)). That is, in the descending branch the bottom FM-layer reverses prior to the top one, whereas in the ascending branch the top FM-layer reverses prior to the bottom one. Moreover, in the case of LNiPy a complete separation of the loops into the top and bottom subloops does not occur.

As it can be seen (Fig. 3 (a), (b)), the slope of hysteresis loops for both LNiPy and HNiPy samples at 2 nm of AFM-layer is smaller than the one at 4 nm. Considering the peaks of the differential susceptibility distribution for descending and ascending branches (Fig. 4), it should be mentioned that the peaks width widens with the increase of the AFM-layer thickness. This means that the contribution to the reversible switching of the magnetization, i.e. magnetic moments rotation, increases with increasing the AFM-layer thickness.

The dependences of estimated values of the exchange bias and coercivity on the thickness of the antiferromagnetic layer for the top and bottom FM-layers are presented in Fig. 5. A decrease in temperature leads to an increase in the exchange bias. It can be explained by the decreasing of the energy of thermal fluctuations in the system. This leads to the spin structure at the interfaces to be more stable, and hence the magnetic moments of the ferromagnetic layer need more energy to overcome this barrier. As the temperature decreases to 80 K, the coercive force of the samples increases. For the samples with IrMn-layer thickness of 2 nm, the blocking temperature below which the exchange bias was found is 250 K for the LNiPy and 200 K for the HNiPy sample. For the samples with the antiferromagnetic layer thickness of 4 nm, the exchange bias was observed at temperatures below 290 K. The tendency to an increase of the blocking temperature with increasing AFM-layer thickness can be explained by an increase of the anisotropy energy of the AFM layer [6,24,27,36].

Thus, it was shown that the mechanism of magnetization reversal is maintained for the Permalloy of each composition with decreasing temperature. The sequence of magnetization reversals of ferromagnetic layers for structures based on LNiPy in the descending branch is the same as in the ascending branch, whereas it is different for the structures based on HNiPy.

**Conclusions**

The studies of magnetic properties performed in the temperature range from 80 K to 300 K allowed us to determine the blocking temperatures of NiFe/IrMn/NiFe thin-film structures with the antiferromagnetic layer thicknesses of 2 nm and 4 nm, which do not exhibit the exchange bias effect at room temperature. The sequence of magnetization reversal for the two ferromagnetic layers has been determined to depend on the AFM-layer thickness and the temperature. Thus, at 4 nm thickness of IrMn layers for the LNiPy samples the top FM-layer reverses prior to the bottom one. On the contrary, for the HNiPy samples the bottom FM-layer reverses prior to the top one. The aforementioned differences are supposedly caused by the structural qualities of the systems based on the LNiPy and HNiPy.

**Acknowledgments**

Ch. Gritsenko and M. Gorshenkov acknowledge support by the Russian Foundation for Basic Research (RFFI grant № 17-32-50170). O.A.T. acknowledges support by the Grants-in-Aid for Scientific Research (Grant Nos. 17K05511 and 17H05173) from MEXT, Japan, by the grant of the Center for Science and Innovation in Spintronics (Core Research Cluster), Tohoku University, and by JSPS and RFBR under the Japan-Russian Research Cooperative Program.

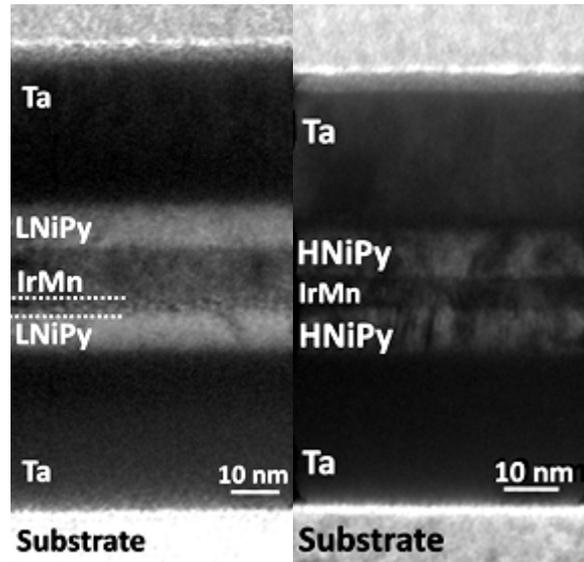

Fig. 1. TEM-images for the samples LNiPy(10 nm)/IrMn(10 nm)/LNiPy(10 nm) and HNiPy(10 nm)/IrMn(5 nm)/HNiPy(10 nm).

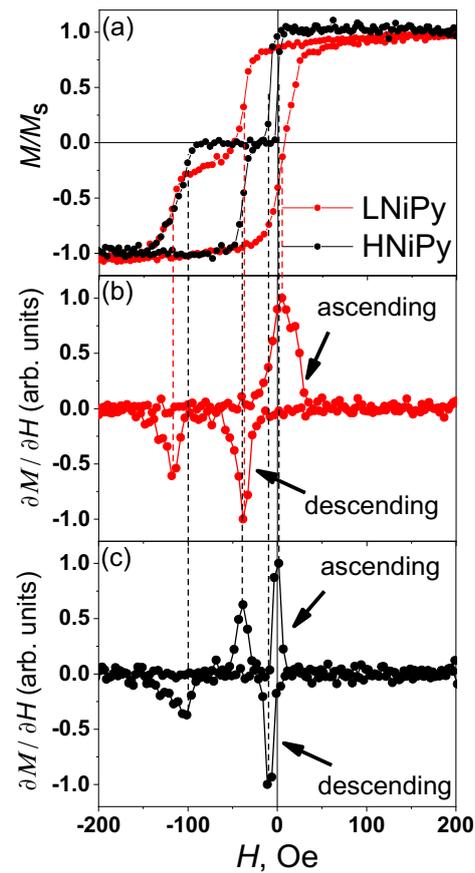

Fig. 2. (a) Hysteresis loops for the LNiPy and HNiPy samples NiFe/IrMn(10 nm)/NiFe, obtained along the unidirectional anisotropy, and corresponding differential susceptibility distribution for the LNiFe (b) and HNiPy (c) samples.

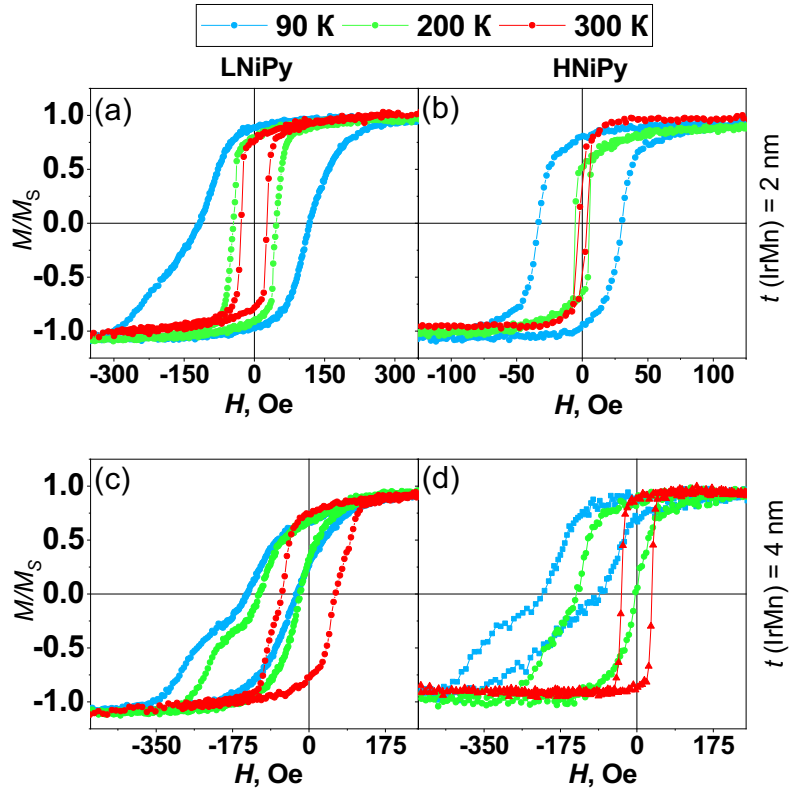

Fig. 3. Hysteresis loops obtained along the unidirectional anisotropy for the LNiPy (a), (c) and HNiPy(b), (d) NiFe/IrMn(t)/NiFe samples.

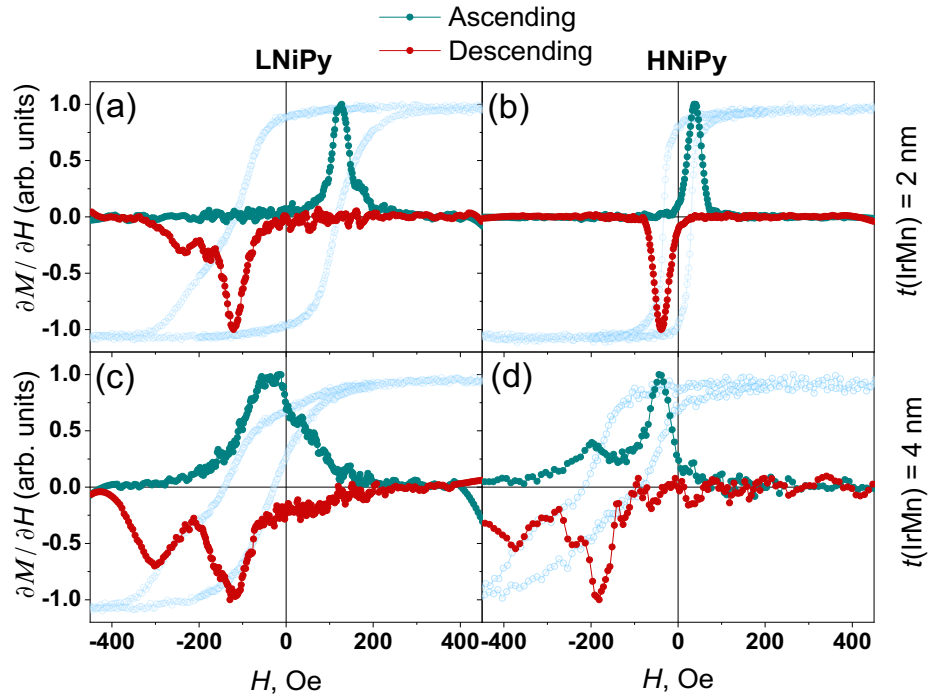

Fig. 4. The differential susceptibility distribution for LNiPy (a), (c) and HNiPy(b), (d) NiFe/IrMn(t)/NiFe samples at 90 K.

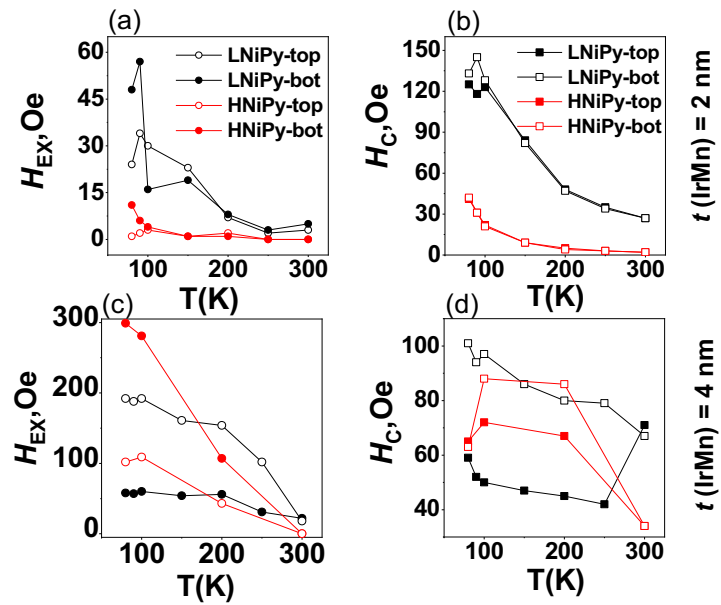

Fig. 5. Exchange bias and coercivity dependences on the temperature for the top and bottom FM-layers in samples NiFe/IrMn(t)/NiFe with LNiPy (a), (c) and HNiPy (b), (d).